\documentclass[a4paper]{article}
\begin{document}
\tolerance=5000
\def\pp{{\, \mid \hskip -1.5mm =}}
\def\cL{{\cal L}}
\def\hs{{\qquad\qquad}}
\def\beq{\begin{eqnarray}}
\def\eeq{\end{eqnarray}}
\def\tr{{\rm tr}\, }
\def\ii{\infty}
\def\Tr{{\rm Tr}\, }
\def\nn{\nonumber}
\def\e{{\rm e}}
\def\D{{D \hskip -3mm /\,}}
\def\maketitle{\thispagestyle{empty}\setcounter{page}0\newpage
                \renewcommand{\thefootnote}{\arabic{footnote}}
                  \setcounter{footnote}0}
\renewcommand{\thanks}[1]{\renewcommand{\thefootnote}{\fnsymbol{footnote}}
               \footnote{#1}\renewcommand{\thefootnote}{\arabic{footnote}}}
\newcommand{\preprint}[1]{\hfill{\sl preprint - #1}\par\bigskip\par\rm}
\renewcommand{\title}[1]{\begin{center}\Large\bf #1\end{center}\rm\par\bigskip}
\renewcommand{\author}[1]{\begin{center}\Large #1\end{center}}
\newcommand{\address}[1]{\begin{center}\large #1\end{center}}
\newcommand{\pacs}[1]{\smallskip\noindent{\sl PACS numbers:
                       \hspace{0.3cm}#1}\par\bigskip\rm}
\renewcommand{\date}[1]{\par\bigskip\par\sl\hfill #1\par\medskip\par\rm}
\def\babs{\hrule\par\begin{description}\item{Abstract: }\it}
\def\eabs{\par\end{description}\hrule\par\medskip\rm}
\newcommand{\guido}{Guido Cognola\thanks{e-mail: \sl cognola@science.unitn.it\rm}}
\newcommand{\sergio}{Sergio Zerbini\thanks{e-mail: \sl zerbini@science.unitn.it\rm}}
\newcommand{\emilio}{Emilio Elizalde\thanks{e-mail: \sl elizalde@ieec.fcr.es\rm\rm}}
\def\dinfn{Dipartimento di Fisica, Universit\`a di Trento\\
                           and Istituto Nazionale di Fisica Nucleare,\\
                                   Gruppo Collegato di Trento, Italia \medskip}

\def\csic{Consejo Superior de Investigaciones Cient\'{\i}ficas\\
                        IEEC, Edifici Nexus 201,
                            Gran Capit\`a 2-4, 08034 Barcelona, Spain\\
                               and Departament ECM and IFAE,
                                 Facultat de F\'{\i}sica,\\
                                      Universitat de Barcelona,
                                 Diagonal 647, 08028 Barcelona, Spain \medskip}


\title{\LARGE One-loop effective potential \\ from higher-dimensional
 AdS black holes}

\author{\guido${}^a$, \emilio${}^b$, \sergio${}^a$}
\address{${}^a$\dinfn \\  ${}^b$\csic}

\vspace*{10mm}

\abstract{We study the quantum effects in a brane-world model in which a
positive constant curvature  brane universe is  embedded
in a higher-dimensional
bulk AdS black hole, instead of the usual portion of the  AdS$_5$.
By using zeta regularisation, in the large mass regime,
we explicitly calculate the
one-loop effective potential due to the bulk quantum fields and show
that it leads to a non-vanishing cosmological constant, which can
definitely acquire a positive value.}
\bigskip

\date{January 2004}

\vspace*{15mm}

\section{Introduction}

The study of quantum effects in the brane-world has been a subject
of quite some activity recently. In particular, the one-loop
effective potential for bulk quantum fields has been calculated,
for the case  when the bulk space is the 5-dimensional AdS and the
brane is flat \cite{Garriga:2001ar}. This is one of the directions
which relates the AdS/CFT correspondence \cite{AdS} with the
brane-world paradigms \cite{Randall:1999vf}. (It is interesting to
note that the one-loop potential obtained from the bulk space may
be useful in the development of the holographic renormalization
group (RG) in the AdS/CFT set up \cite{BK}-\cite{noji00h}.) Among the
applications of a calculation of this type, one can envisage
radion  stabilisation and the derivation of an induced
cosmological constant. Subsequently, this kind of calculation has
been generalised to the case when the brane is de Sitter space
\cite{Nojiri:2000bz}-\cite{Moss:2003zk}, which is an
interesting example since this setup seems to be the one
corresponding to our observable universe. The main technical
ingredient in our calculation, which allows to carry it explicitly
to the end, will be zeta-function regularisation
\cite{eliz94b,byts96}.

In view of the recent developments on cosmological models, it seems
interesting to generalise the study of quantum effects due
to bulk fields to different sorts of bulk spaces. Indeed, if our
universe is a brane-world of any kind, it is unclear a priori what is
the right bulk space. For example, it is already
known that brane gravity trapping occurs in an AdS 5-dimensional black hole
\cite{gomez}-\cite{Nojiri:2002td} in just
the same way as in the Randall-Sundrum model \cite{Randall:1999vf}.
Moreover, in the AdS/CFT
correspondence, the case of a bulk AdS black hole represents a different
phase of the same theory and there is the exciting connection that a
transition between an ordinary bulk AdS and a bulk  AdS black  hole
corresponds to the
confinement-deconfinement transition in the dual CFT \cite{Witten:2001ua}.
Thus, it appears quite natural to expect that,  at some epoch or other of
its evolution,  our brane universe may have been  embedded into a bulk AdS
black hole.

The purpose of this paper is ---by the way of a calculable explicit example---
 to give  consistency to the ideas
 above. With this aim, we consider the physical de Sitter brane universe to be
 embedded into a higher-dimensional AdS black hole. We show that, by
 using zeta-regularisation techniques, one can explicitly calculate the
one-loop effective potential due to the bulk quantum fields. For
the sake of simplicity, a bulk scalar will be here considered, but
the calculation could be easily extended to other cases. The
one-loop effective potential that we shall obtain will then be
compared with the potential found in the case when the bulk is a
pure AdS space. Such one-loop effective potential may be actually
responsible for the dominant contribution to the brane
cosmological constant during some period of the evolution of our
brane-world. One should note on passing  that applications of the
one-loop effective potential from a bulk AdS black hole to
inducing the correct hierarchy does not look so interesting, since
such kind of  background does not seem at present to be able to
provide a
 natural solution of this problem \cite{Nojiri:2002pf}.

To begin with, as a background bulk space, we may consider an
asymptotically AdS generic black hole  solution, in $D=2+d$
dimension, with Euclidean time $\tau$ and extra radial coordinate
$r$. The metric reads \cite{peldan}-\cite{birm} \beq \label{III}
ds^2=g_{MN}dx^Mdx^N = A(r)d\tau^2 + \frac{1}{A(r)}dr^2 +r^2
d\Sigma^2_{d}, \eeq where $M,N=1,...,D$, $\Sigma_d$ is the
constant curvature  brane space-time and \beq
A(r)=k+\frac{r^2}{\ell^2}-\frac{r_0^{d+1}}{\ell^2r^{d-1}}\:. \eeq
Here $k=0,1,-1$ for Minkowski, De Sitter and anti De Sitter space,
 respectively,
$\ell$ is related to the bulk cosmological constant, and $r_0$ is
a typical length parameter, which depends on the mass of the black
hole and on the bulk Newton constant. In the non extremal case,
the function $A(r)$ has a simple zero at $r=r_H$, which is the
minimum (positive) value admisible for $r$. For the particular
case $k=0$, one trivially has $r_H=r_0$.

\section{Near-horizon approximation}

As is well known,  one-loop calculations in a generic black-hole
background are hard, if not impossible, to perform exactly. One is
compelled to make use of some approximation. Here, we will
investigate quantum  one-loop effects in the so called
near-horizon approximation. This one proves to be a good
approximation to the exact problem in hand as far as the black
hole mass ---in our case the length parameter $r_H$--- is
sufficiently large. In this situation the bulk black hole
space-time becomes a Rindler-like space-time and the statistical
mechanics of the black holes can be investigated in detail (for a
further discussion see, for example \cite{zerbini}).

In fact, defining the new coordinates, $\rho$ and $\theta$, by
means of \beq r=r_H+\frac{A'(r_H)}{4}\:\rho^2\,,\qquad\qquad
\tau=\frac2{A'(r_H)}\:\theta\:, \eeq one gets \beq ds^2 \sim
\rho^2d\theta^2 + d\rho^2 + r_H^2 d\Sigma^2_d\,. \eeq As usual, in
order to avoid the conical singularity at $\rho=0$, one has to
require the coordinate $\theta$ to be periodic with a period of
$2\pi$. This corresponds to a period $\beta$ of the Euclidean time
$\tau$ given by the Hawking condition \beq
\beta=\frac{4\pi}{A'(r_H)}\,. \eeq In this way, the space-time
becomes locally $R_2\times\Sigma_d$, this is to say,  the
Euclidean version of a Rindler-like space-time.

Consider then the action for a scalar with scalar-gravitational
coupling in the bulk, e.g. \beq \label{act3} {\cal S} = {1\over
2}\int d^Dx\:\sqrt{g}\: \left[ g^{M N}\partial_{M} \phi
\partial_{N} \phi +m^2 \phi ^2 + \xi R \phi^2 \right] \;, \eeq $m$
being the mass, $\xi$ the constant coupling with gravitation, and
$R$ the scalar curvature of the whole manifold. The latter action
can be rewritten as \beq \label{LC5b} {\cal S} = {1\over 2}\int
d^Dx\:\sqrt{g}\:\phi L \phi\ , \qquad\qquad
L=L_D+M^2=L_2+L_d+M^2\,, \eeq where $L_D=L_2+L_d$ is a
Laplacian-like operator on the $D$-dimensional manifold, while
\beq L_2\phi \equiv -\frac1{\rho^2}\partial_\theta^2\phi -
\frac1\rho\partial_\rho\left(\rho\:\partial_\rho\phi\right) \eeq
is the flat Laplacian in 2 dimensions, \beq L_d\phi\equiv
\left[-\nabla^2_d+\frac{(d-1)^2}{4r_H^2}\right]\phi \label{e9}\eeq
is a Laplacian-like operator (with a particular  non mimimal
coupling) on the constant curvature space-time $\Sigma_d$ and,
finally, \beq M^2=m^2+\xi R-\frac{(d-1)^2}{4r_H^2}=
m^2+\frac{1}{r_H^2}\left[\xi\:d(d-1)-\frac{(d-1)^2}{4}\right]
\label{e10} \eeq is a constant term. For computational reason, we
have added and subtracted in expressions (\ref{e9}) and
(\ref{e10}) the constant term $(d-1)^2/(4r_H^2)$.

\section{One-loop effective potential}

Since we are interested in the effective potential, we have to
compute the heat-kernel trace and then, via the Mellin transform,
the zeta function corresponding to the operator $L$. In fact, in
the zeta-function regularisation scheme, the one-loop contribution
to the effective potential is given by \beq
V^{(1)}=-\frac{\zeta'(0|L/\mu^2)}{2V_D}=
-\frac{\zeta'(0|L)+\log\mu^2\: \zeta(0|L)}{2V_D}\:, \eeq $V_D$
being the volume of the manifold and $\mu$ a free parameter, which
one has to introduce for dimensional reasons. It must be fixed by
renormalization. Following Ref.~\cite{cher}, we write the
effective potential in the form \beq
V_{eff}&=&V_r(\mu)+V^{(1)}(\mu)\:, \eeq where $V_r(\mu)$ is the
renormalized vacuum energy. The effective potential is a physical
observable and for this reason it cannot depend on the choice of
the arbitrary scale parameter $\mu$. This means that it has to
satisfy the renormalization condition \cite{cher} \beq
\mu\:\frac{dV_{eff}}{d\mu}=0\:. \eeq From the latter equation we
determine $V_r(\mu)$ up to an integration constant $V_r(\mu_0)$,
which we choose to be vanishing, and thus obtain in this way the
renormalization point $\mu_0$. After such an operation one finally
gets the renormalized effective potential in the form \beq
V_{eff}=-\frac{\zeta'(0|L/\mu_0^2)}{2V_D}=
-\frac{\zeta'(0|L)+\log\mu_0^2\: \zeta(0|L)}{2V_D}\:,
\label{veff}\eeq

In the approximation we are considering, $L_2$ and $L_d$ commute,
thus (here $t$ is the heat-kernel parameter) \beq \label{LC6} \Tr
e^{-t L}=\mbox{Tr}e^{-t L_2}\Tr e^{-t L_d}e^{-tM^2}\:. \eeq Let us
put the branes at say $\rho=\rho_1$ and $\rho=\rho_2$ (with
$\rho_1>\rho_2$). Then the eigenfunctions and the eigenvalues of
$L_2$ are given by \beq \label{LC7} && -L_2 \phi = \lambda^2 \phi\
,\quad \quad \phi = \e^{in\theta}\left(\alpha_n J_n(\lambda\rho)
+ \beta_n N_n(\lambda \rho)\right)\ ,\nn\\
&& \lambda = \lambda_n\ , \quad\quad
\left(n=0,\pm1,\pm2,\cdots\right)\ . \eeq Here $J_n$ and $N_n$ are
 Bessel and the Neumann functions, respectively. If, for
simplicity, we impose Dirichlet boundary condition at the branes,
that is \beq \label{LC8} \phi(\rho_1)=\phi(\rho_2)=0\ , \eeq then,
the eigenvalues $\lambda_n$ are implicitly given by the equation
\beq \label{LC9} J_n\left(\lambda_n \rho_1\right)
N_n\left(\lambda_n \rho_2\right) =J_n\left(\lambda_n \rho_2\right)
N_n\left(\lambda_n \rho_1\right)\ . \eeq

For simplicity, now let us compute the effective potential for the
scalar field in the bulk with olny one brane. In fact, it is not
difficult to realize that the two-brane case does not add any
additional physical insight (although the calculation is rather
more involved). We put the brane at $\rho=\rho_0$, the bulk space
being defined by $\rho<\rho_0$. Thus, it turns out that the
eigenfunctions and eigenvalues of $L_2$ are given by \beq
\label{LC10} && L_2 \phi = \lambda^2 \phi\ ,\quad
\phi = \e^{in\theta} J_n(\lambda \rho) \ ,\nn\\
&& \lambda = \lambda_{n,k}={j_{n,k} \over \rho_0} \ , \quad
\left(n=0,\pm1,\pm2,\cdots\right)\ . \eeq Here $j_{n,k}$ are the
zeros  of the Bessel functions $J_n(x)$:
$J_n\left(j_{n,k}\right)=0$.

\section{Large mass expansion}

As it always happens in a situation of this kind, it is not
possible to do the calculation exactly in a closed way, valid in
the whole range of parameters of the problem. For different
domains of the parameters a different expansion must be chosen.
Here, we choose to perform an expansion for large values of the
constant term $M^2$ (with respect to the renormalization point
$\mu_0^2$)\footnote{Which will fix, all the way from now on, the
unit in which $M^2$ is to be measured.}, since this one
corresponds to the most interesting (and natural) situation from
the physical viewpoint. In this situation the zeta function
acquires the form \beq \zeta(s|L)\sim\sum_{r=0}^{\infty} K_r(L_D)
\frac{\Gamma(s+\frac{r-D}{2})}{\Gamma(s)}M^{D-r-2s}\,. \eeq The
latter expression directly follows from the Mellin-like transform
\beq \zeta(s)=\frac{1}{\Gamma(s)}\int_{0}^{\infty}\:
dt\:t^{s-1}\Tr e^{-tL}\:, \label{mlt1}\eeq by using the
heat-kernel expansion \beq \Tr
e^{-tL}\sim\sum_{r=0}^{\ii}\:e^{-tM^2}K_r(L_D)t^{\frac{r-D}2}\:,
\label{hke1}\eeq $K_r(L_D)$ being the Seeley-DeWitt coefficients
corresponding to the operator $L_D$.

A direct computation gives
\beq
\zeta(0|L)=K_D(L)=\sum_{n\leq D\:\:even}
\:\frac{(-1)^{n/2}K_{D-n}(L_D)M^{n}}{\left(\frac{n}2\right)!}\:,
\label{zeta0}\eeq
while
\beq
\zeta'(0|L)&=& \sum_{n\leq D\:\: even}
\frac{(-1)^{n/2}K_{D-n}(L_D)M^n}{\left(\frac{n}2\right)!}
\:\left[-\log M^2+\gamma+\psi(1+n/2)\right]
\nonumber \\ &&
+\sum_{n\leq D\:\:odd}
\Gamma(-n/2)K_{D-n}(L_D)M^n
+O\left(\frac1M\right)\:,
\label{zeta1}\eeq
$\gamma$ being the Euler-Mascheroni constant and $\psi$ the digamma function
(logarithmic derivative of the $\Gamma$ function).

The heat-kernel coefficients $K_n(L_D)$ can be  given in terms of
$K_i(L_2)$ and $K_j(L_d)$. In fact, since \beq \label{bbb1} \Tr
e^{-tL_D}=\Tr e^{-tL_2}\Tr e^{-tL_d}
\sim\sum_{n=0}K_n(L_D)t^{\frac{n-D}{2}} \eeq and \beq \label{bbb2}
\Tr e^{-tL_2}\sim\sum_{i=0}K_i(L_2)t^{\frac{i-2}{2}}\:,
\qquad\qquad \label{bbb3} \Tr
e^{-tL_d}=\sum_{j=0}A_j(L_d)t^{j-d/2}\:, \eeq one gets \beq
K_n(L_D)=\sum_{i+2j=n}K_i(L_2)A_j(L_d)\,. \label{knD} \eeq Here we
have used the notation $A_j=K_{2j}$ since on a compact manifold
without boundary, $\Sigma_d$, all $K_j$ coefficients with odd
index vanish. The $A_n(L_d)$ coefficients depend on the horizon
manifold and, in principle, they can be computed in terms of
geometric invariants. The other coefficients $K_n(L_2)$ are
associated with the Dirichlet Laplacian on the disk with radius
$\rho_0$. For dimensional reasons,\footnote{Note that the
dimensions of the $R_i$ take care of the corresponding powers of
$M$.} they have the form \beq K_i(L_2)=d_i\rho_0^{2-i}\,,
\label{kn2} \eeq where $d_i$ are numerical coefficients which can
be evaluated with the help of the techniques developed in Refs.
\cite{bord96-37}. The explicit values of the ones we shall use in
the following read \beq && d_0=\frac{1}{4}\,,\qquad
d_2=\frac{5}{12}\,,\qquad d_4=\frac{347}{5040}\,,\qquad
d_6=\frac{602993}{5765760}\,, \nn\\&&
d_1=-\frac{\sqrt{\pi}}{4}\,,\qquad
d_3=\frac{\sqrt{\pi}}{128}\,,\qquad d_5=\frac{25}{192
\sqrt{\pi}}+\frac{37\sqrt{\pi}}{16384}\,. \eeq

Eqs. (\ref{zeta0}) and (\ref{zeta1}) are valid in arbitrary
dimensions. For cosmological applications, however, it is
interesting to specify for the $D=6$ case, which corresponds to a
4-dimensional brane as horizon manifold. From Eqs.~(\ref{veff}),
(\ref{zeta0}) and (\ref{zeta1}), for $D=6$ and in the large $M$
limit, we obtain \beq V_{eff}&=&\frac{1}{2V_D}\left[
\frac{8\sqrt\pi}{15}K_1(L_6)M^5 -\frac{4\sqrt\pi}{3}K_3(L_6)M^3
+2\sqrt\pi K_5(L_6)M \right.\nonumber \\ &&\qquad
-\frac{K_0(L_6)M^6}{6}\:\left(\log\frac{M^2}{\mu_0^2}-\frac{11}6\right)
+\frac{K_2(L_6)M^4}{2}\:\left(\log\frac{M^2}{\mu_0^2}-\frac32\right)
\nonumber \\ &&\qquad\qquad\left.
-K_4(L_6)M^2\:\left(\log\frac{M^2}{\mu_0^2}-1\right)
+K_6(L_6)\:\log\frac{M^2}{\mu_0^2}\right]+O\left(\frac1M\right)\:.
\label{veff6}\eeq

In the following we shall analyze in detail the case $k=1$, that
is, the de Sitter brane, which is most promising from the
cosmological viewpoint. The simplest case $k=0$, that is the
Minkowski brane, will be directly obtained in the limit of
vanishing curvature.

\section{The de Sitter brane case}

As anticipated, the de Sitter case ($k=1$) may be very interesting
for cosmological applications. The results that we obtain for this
case are the following.

The heat-kernel coefficients for the operator $-\nabla^2+9/4r_H^2$
on the 4-dimensional sphere can be taken, for instance, from
Ref.~\cite{byts96}. The non-vanishing ones read \beq
A_0=\frac{V_4}{16\pi^2}\:,\hs A_1=-\frac{V_4}{64\pi^2r_H^2}\:,\hs
A_2=\frac{V_4}{16\pi^2r_H^4}\:. \eeq From Eqs.~(\ref{knD})
and (\ref{kn2}) and setting $V_6=\pi\rho_0^2V_4$, it follows that
\beq K_n(L_6)=\frac{V_6}{16\pi^3}\left( \frac{d_n}{\rho_0^n}
-\frac{d_{n-2}}{4r_H^2\rho_0^{n-2}}
+\frac{d_{n-4}}{r_H^4\rho_0^{n-4}}\right)\:.\eeq Now,
using Eq.~(\ref{veff6}) we are able to write the final result
under the explicit form \beq V_{eff}&=&\frac{1}{\rho_0^6}\left[
\frac{d_1(M\rho_0)^5}{60\pi^{5/2}}
-\frac{(M\rho_0)^3}{24\pi^{5/2}}
\left(d_3-\frac{d_1\rho_0^2}{4r_H^2}\right)
+\frac{M\rho_0}{16\pi^{5/2}}\left(d_5-\frac{d_3\rho_0^2}{4r_H^2}
+\frac{d_1\rho_0^4}{r_H^4}\right)\right. \nonumber\\ && -
\frac{d_0(M\rho_0)^6}{192\pi^3}
      \:\left(\log\frac{M^2}{\mu_0^2}-\frac{11}{6}\right)
+\frac{(M\rho_0)^4}{64\pi^3}\left(d_2-\frac{d_0\rho_0^2}{4r_H^2}\right)
     \:\left(\log\frac{M^2}{\mu_0^2}-\frac{3}{2}\right)
\nonumber \\ &&\qquad\qquad\qquad\qquad
-\frac{(M\rho_0)^2}{32\pi^3}\left(d_4-\frac{d_2\rho_0^2}{4r_H^2}
     +\frac{d_0\rho_0^4}{r_H^4}\right)
     \:\left(\log\frac{M^2}{\mu_0^2}-1\right)
\nonumber \\ &&\qquad\qquad\qquad\qquad\qquad\left.
+\frac{1}{32\pi^3}\left(d_6-\frac{d_4\rho_0^2}{4r_H^2}
     +\frac{d_2\rho_0^4}{r_H^4}\right)
     \:\log\frac{M^2}{\mu_0^2}
\right]+O\left(\frac1M\right)\:. \label{zeta1-6}\eeq

\section{The flat  brane case}
The one-loop effective potential
for the simplest case $k=0$ (Minkowski)
can be obtained from Eq.~(\ref{veff6}) by observing that
\beq
K_n(L_6)=\frac{\rho_0^2V_4d_n}{16\pi^2\rho_0^n}=
\frac{V_6d_n}{16\pi^3\rho_0^n}\,,
\eeq
or, more simply, it can be obtained by taking the flat limit
$r_H\to\infty$ in Eq.~(\ref{zeta1-6}). We get
\beq
V_{eff}&=&\frac{1}{\rho_0^6}\left[
\frac{d_1(M\rho_0)^5}{60\pi^{5/2}}
-\frac{d_3(M\rho_0)^3}{24\pi^{5/2}}
+\frac{d_5M\rho_0}{16\pi^{5/2}}
\right. \nonumber\\ &&
- \frac{d_0(M\rho_0)^6}{192\pi^3}
      \:\left(\log\frac{M^2}{\mu_0^2}-\frac{11}{6}\right)
+\frac{d_2(M\rho_0)^4}{64\pi^3}
     \:\left(\log\frac{M^2}{\mu_0^2}-\frac{3}{2}\right)
\nonumber \\ &&\qquad\qquad\left.
-\frac{d_4(M\rho_0)^2}{32\pi^3}
     \:\left(\log\frac{M^2}{\mu_0^2}-1\right)
+\frac{d_6}{32\pi^3}
     \:\log\frac{M^2}{\mu_0^2}
\right]+O\left(\frac1M\right)\:. \label{efp1}\eeq

\section{Concluding remarks}
Performing a numerical analysis of the result above, we arrive to
the conclusion that such quantity ---which should be identified
essentially with the cosmological constant--- is generically
non-zero and can acquire positive and  negative values, depending
on the specific choice of the parameters. The same is seen to
happen for a dS brane in the bulk we are considering, Eq.
(\ref{zeta1-6}). This has now to be compared with the Casimir
effect for a dS brane in the AdS bulk: in that case the
cosmological constant is always zero \cite{Nojiri:2000bz}. A more
detailed analysis shows actually, that for the range of values for
$M$ and $\rho_0$ of physical interest, which have been under
discussion  in the recent literature: (i) all the series we have
here converge very quickly (this already happens for $\rho_0 >
10^{-10}$ cm), and (ii) the value of the induced cosmological
constant that we obtain from them is positive, as needed to
explain the observed acceleration in the expansion of the
universe. In this preliminary analysis it is too soon to discuss
about a numerical matching with the observational
values.\footnote{See e.g. \cite{pad1} for alternative and
complementary approaches.} Similarly, it can be shown from Eq.
(\ref{zeta1-6}), that the effective potential corresponding to the
case of a de Sitter brane is also non-zero, and can be made
positive too, providing an enlarged number of interesting
situations.

Summing up, from these simple examples we have here considered it
becomes already clear that, in order to induce a 4-dimensional
cosmological constant in a brane-world universe, it turns out that
an AdS black-hole bulk with a one-brane configuration is far more
attractive than the pure AdS bulk. At this stage we have dealt only with
some simple examples but, as has been pointed out already, the
present calculation can be extended, without essential trouble, to
more realistic situations and, thus, it seems to open a number of
different, very promising possibilities.

\section*{Acknowledgements}
We thank S.~D.~Odintsov for helpful discussions and suggestions.
This investigation has been supported by the Program
INFN(Italy)-CICYT(Spain), and by DGICYT (Spain), project BFM2000-0810.


\begin{thebibliography}{99}

\bibitem{Garriga:2001ar}
J.~Garriga, O.~Pujolas and T.~Tanaka,
Nucl.~Phys. {\bf B605} (2001) 192; arXiv:hep-th/004109, hep-th/0111277;
I.~Brevik, K.~A.~Milton, S.~Nojiri and S.~D.~Odintsov,
Nucl. Phys.  {\bf B599} (2001) 305 [arXiv:hep-th/0010205];
R.~Hofmann, P.~Kanti and M.~Pospelov,
Phys. Rev.  {\bf D63} (2001) 124020 [arXiv:hep-ph/0012213];
A.~Flachi, I.~G.~Moss and D.~J.~Toms,
Phys. Rev.  {\bf D64} (2001) 105029. [arXiv:hep-th/0106076];
A.~A.~Saharian and M.~R.~Setare,
Phys. Lett.  {\bf B552} (2003) 119 [arXiv:hep-th/0207138];
S.~Nojiri and S.~D.~Odintsov, arXiv:hep-th/0302054;
A.~Flachi, J.~Garriga, O.~Pujolas and T.~Tanaka,  JHEP {\bf 0308} (2003) 053
[arXiv:hep-th/0302017];
S. Nojiri, S.~D. Odintsov and S. Ogushi,
Phys. Lett. B {\bf 506} (2001) 200 [arXiv:hep-th/0102082].

\bibitem{AdS} J.~M.~Maldacena,  Adv. Theor. Math. Phys. {\bf 2}
(1998) 231; E. Witten,  Adv. Theor. Math. Phys. {\bf 2} (1998) 253;
S. Gubser, I.~R. Klebanov and A.~M. Polyakov,  Phys. Lett. {\bf B428}
(1998) 105.

\bibitem{Randall:1999vf}
L.~Randall and R.~Sundrum,
Phys. Rev. Lett.  {\bf 83} (1999) 4690
[arXiv:hep-th/9906064].


\bibitem{BK} V. Balasubramanian and P. Kraus,
 Phys. Rev. Lett. {\bf 83} (1999) 3605 [arXiv:hep-th/9903190].

\bibitem{VV} E. Verlinde and H. Verlinde, ``RG-flow, gravity and the
cosmological constant,''  arXiv:hep-th/9912018.

\bibitem{noji00h} S.~Nojiri, S.~D.~Odintsov and S.~Zerbini,
 Phys. Rev.  {\bf D62} (2000) 064006 [arXiv:hep-th/0001192].


\bibitem{Nojiri:2000bz}
S.~Nojiri, S.~D.~Odintsov and S.~Zerbini,
Class. Quant. Grav.  {\bf 17} (2000) 4855
[arXiv:hep-th/0006115].

\bibitem{Elizalde:2002dd}
E.~Elizalde, S.~Nojiri, S.~D.~Odintsov and S.~Ogushi,
Phys. Rev. {\bf D67} (2003) 063515 [arXiv:hep-th/0209242];
W.~Naylor and M.~Sasaki,
Phys. Lett.  {\bf B542} (2002) 289
[arXiv:hep-th/0205277].

\bibitem{Moss:2003zk}
I.~G.~Moss, W.~Naylor, W.~Santiago-German and M.~Sasaki, Phys.
Rev.  {\bf D67} (2003) 125010 [arXiv:hep-th/0302143].

\bibitem{eliz94b}
E.~Elizalde, S.~D.~Odintsov, A.~Romeo, A. A.~Bytsenko and
S.~Zerbini. {\em Zeta Regularisation Techniques with
Applications}. (World Scientific, Singapore,  1994); E. Elizalde,
{\em Ten physical applications of spectral zeta functions}
(Springer, Berlin, 1995).

\bibitem{byts96}
A.A.~Bytsenko, G.~Cognola, L.~Vanzo and S.~Zerbini.
Phys. Rept.  {\bf 266} (1996) 1
[arXiv:hep-th/9505061].

\bibitem{gomez}
 C. ~Gomez, B. ~Janssen, and P.J. ~Silva.
 JHEP {\bf 0004} (2000) 027.
[arXiv:hep-th/0003002].


\bibitem{moschella}
A.~Kamenshchik, U.~Moschella, V.~Pasquier.
 Phys. Lett. {\bf B487} (2000) 7
[arXiv:gr-qc/0005011].


\bibitem{rinaldi}
D.~Birmingham and M.~Rinaldi.
 Mod. Phys. Lett. {\bf A16} (2001) 1887
[arXiv:hep-th/0106237].

\bibitem{Nojiri:2002td}
S.~Nojiri and S.~D.~Odintsov,
 Int. J. Math. Phys.  {\bf A18} (2003) 2001 [arXiv:hep-th/0211023].

\bibitem{Witten:2001ua}
E.~Witten. Adv. Theor. Math. Phys. {\bf 2} (1988) 505
[arXiv:hep-th/0112258].

\bibitem{Nojiri:2002pf}
S.~Nojiri and S.~D.~Odintsov,
 Phys. Lett.  {\bf B562} (2003) 9 [arXiv:hep-th/0212306].

\bibitem{peldan}
S. Aminneborg, I. Bengtsson, S. Holst and P. Peldan,
Class. Quant. Grav. {\bf 13} (1996) 2707.

\bibitem{mann}
R. B. Mann, Class. Quant. Grav.  {\bf 16} (1997) L109.
\bibitem{vanzo}
L. Vanzo, Phys. Rev.  {\bf D56} (1997) 6475.
\bibitem{birm}
D. Birmingham, Class. Quant. Grav.  {\bf 16} (1999) 1197.

\bibitem{zerbini}
S. Zerbini, G. Cognola and L. Vanzo.
 Phys. Rev.  {\bf D54} (1996) 2699.

\bibitem{cher}
I. O. Cherednikov, Acta Phys.Slov. {\bf 52} (2002) 221
[arXiv:hep-th/0206245].

\bibitem{bord96-37}
M. Bordag, E. Elizalde, K. Kirsten,  J. Math. Phys. {\bf 37} (1996) 895;
E. Elizalde, M. Bordag and K. Kirsten,  J. Phys. {\bf A31}
 (1998) 1743; M. Bordag, E. Elizalde, K. Kirsten and S. Leseduarte,
 Phys. Rev. {\bf D56} (1997) 4896; M. Bordag, E. Elizalde, B. Geyer
and K. Kirsten,  Commun. Math. Phys. {\bf 179} (1996) 215; G.
Cognola, E. Elizalde and K. Kirsten,  J. Phys. {\bf A34} (2001)
7311.

\bibitem{pad1} T. Padmanabhan, Phys. Rept. {\bf 380} (2003) 235;
 T. Padmanabhan, Class. Quant. Grav.  {\bf 19} (2002) L167;
E. Elizalde,  Phys. Lett.  {\bf B516} (2001) 143; arXiv:hep-th/0311195.

\end{thebibliography}
\end{document}